\documentclass[elsarticle,floatfix,showpacs,showkeys,epsfig,graphics]{revtex4-1}%

\usepackage{graphicx}
\usepackage{amsmath}
\usepackage{amsfonts}
\usepackage{amssymb}
\usepackage{filecontents}

\newcommand{\newc}{\newcommand}
\newc{\beq}    {\begin{equation}}
\newc{\eeq}    {\end{equation}}

\newc{\beqa}    {\begin{eqnarray}}
\newc{\eeqa}    {\end{eqnarray}}
\newc{\bs}    {\section}
\newc{\no}    {\\ \nonumber}

\def\apj{{\em Astrophys. J.  }}
\def\apjl{{\em Astrophys. J. Lett. }}

\def\mnras{{ Mon. Not. Roy. Astron. Soc.  }}

\newc{\st}    {\stackrel}

\begin{document}
\title{Characteristic Size and Mass  of  Galaxies in the Bose-Einstein Condensate Dark Matter Model}
\author{Jae-Weon Lee}\email{scikid@jwu.ac.kr}
\affiliation{Department of Physics, North Carolina State University, Raleigh, NC 27695, USA}
\affiliation{Department of renewable energy, Jungwon university,
            85 Munmu-ro, Goesan-eup, Goesan-gun, Chungcheongbuk-do,
              367-805, Korea}

\begin{abstract}
We study the characteristic  length scale of galactic halos  in the Bose-Einstein condensate (or scalar field) dark matter model.
Considering  the evolution of the density perturbation
we show that the average background  matter density determines the quantum Jeans mass and hence
the spatial size of galaxies at a given epoch. In this model the minimum size of galaxies increases while
the minimum mass of the  galaxies decreases as the universe expands.
The observed values
of the mass and the size of the dwarf galaxies are successfully reproduced with
the dark matter particle mass $m\simeq 5\times 10^{-22}eV$.
The minimum size is about $6\times 10^{-3}\sqrt{m/H}\lambda_c$
and the typical rotation velocity of the dwarf galaxies is $O(\sqrt{H/m}$) c, where $H$ is the Hubble parameter and
$\lambda_c$ is the Compton wave length of the particle.
We also suggest that  ultra compact dwarf galaxies
are the remnants of the dwarf galaxies formed in the early universe.
\end{abstract}

\maketitle


One of the long standing questions in astronomy is what  determines the size of galaxies.
 In this paper we show that the Bose-Einstein
condensate (BEC) dark matter (DM) or the scalar field dark matter (SFDM) can explain the minimum size and the mass of  galaxies in a unified way.

DM remains a great mystery in astrophysics, particle physics and cosmology.
The cold dark matter (CDM) model   is  very successful in explaining the
large scale structures in the universe, but
has many problems in explaining galactic structures.
For example, one of the early evidences for the DM presence is the flatness of galactic
rotation curves  ~\cite{1990PhRvL..64.1084P}, however the CDM is not so successful in explaining the
rotation curves in  galaxy cores.
 Numerical studies with $\Lambda$CDM model predict a cusped halo central  density
 and many subhalos, which are also in discord with  observational data~\cite{Salucci:2002nc,navarro-1996-462,deblok-2002,crisis}.
On the other hand, the BEC/SFDM  ~\cite{1983PhLB..122..221B,1989PhRvA..39.4207M,sin1,myhalo} can be a good alternative to the CDM, because
the BEC/SFDM plays the role of the CDM at  super-galactic scales
and  suppresses sub-galactic structures.
 In this model
 the DM is  a BEC of   the scalar particles with the ultra-light mass $m\simeq 5\times  10^{-22} eV$, whose  quantum nature
   prevents the  formation of the structures smaller than a galaxy  due to the long  Compton wavelength $\lambda_c=2\pi \hbar/mc\simeq 0.08 pc $.

There are two other difficulties the CDM models encounter.
First,
the studies on satellite dwarf spheroidal (dSph)
galaxies of the Milky Way~\cite{gilmore-2008,gilmore-2006}
 indicate that a
typical dSph never has a
size  $< kpc$,
 and that the mass enclosed within the radius
of $300~pc$ in dwarf galaxies is approximately constant ($\sim
10^7 M_\odot$) regardless of their luminosity  ~\cite{Strigari:2008ib}. This result implies the
existence of a  minimum mass scale in addition to the minimum length scale for
 DM dominated objects~\cite{Mateo:1998wg,gilmore-2008}. However,
without introducing the roles of visible matter
 the CDM models usually predict DM dominated structures  down to
$10^{-6}M_\odot$.
Second,
the  observations~\cite{Daddi:2005ym,2009Natur.460..717V,Trujillo21112007} of the size evolution of the most massive galaxies imply
that these galaxies rapidly grow their size  about $5$ times since  $z\sim 2$ while in the CDM models  we expect compact early   galaxies
having  smaller masses.

In Ref. \citealp{Lee:2008jp} we showed that BEC/SFDM can explain the minimum mass of dwarf galaxies, if there is a minimum length scale.
We also proposed that the size evolution of the massive galaxies can be attributed to the evolution of a length scale $\xi$ of
BEC DM~\cite{Lee:2008ux}. In these works we considered the various length scales for $\xi$
such as  $\lambda_c$, a thermal de Broglie wavelength or
a self-interaction scale. For all galaxies $\xi \gg \lambda_c$, and
 we need to find  the exact physical origin of this long scale,
which is the main subject of this paper.

The conjecture that DM is in BEC has a long history.
(See Refs. \citealp{2009JKPS...54.2622L,2014ASSP...38..107S,2014MPLA...2930002R,2014PhRvD..89h4040H,2011PhRvD..84d3531C,2014IJMPA..2950074H}
 for a  review.)
Baldeschi et al.~\cite{1983PhLB..122..221B} studied the galactic halos of
self-gravitating bosons, and Membrado et al.~\cite{1989PhRvA..39.4207M} calculated the rotation curves
of self-gravitating boson halos.
Sin \cite{sin1}  suggested that the halos
are   like  atoms made of ultra-light BEC DM.
 Lee and Koh ~\cite{myhalo} suggested that the  DM halos are the giant boson stars described by the relativistic scalar field theory.
Similar ideas were suggested  by many authors~\cite {1993ApJ...416L..71W,Schunck:1998nq, PhysRevLett.84.3037,PhysRevD.64.123528,repulsive,fuzzy,
 corePeebles,Nontopological,PhysRevD.62.103517,Alcubierre:2001ea,2012PhRvD..86h3535P,2009PhRvL.103k1301S,
 Fuchs:2004xe,Matos:2001ps,0264-9381-18-17-101,PhysRevD.63.125016,Julien,Boehmer:2007um, Matos:2001ps,Eby:2015hsq}.
In literature it has been shown that
BEC/SFDM could explain the  many observed aspects such as rotation curves
~\cite{PhysRevD.64.123528,0264-9381-17-1-102,Mbelek:2004ff,PhysRevD.69.127502},
 the large scale structures of the universe~\cite{2014NatPh..10..496S}, the cosmic background radiation,
 and spiral arms~\cite{2010arXiv1004.4016B}.

 In this paper, we show that BEC/SFDM  has the natural length scale and the mass scale
 determined only by background matter density and the DM particle mass $m$.
In the BEC DM model \cite{sin1}  a galactic DM halo is described with the wave function $\psi(\bold{r})$,
which is the solution of
the Gross-Pitaevskii equation (GPE)
\beq
\label{GPE}
i\hbar\partial_t \psi (\bold{r},t)=-\frac{\hbar^2}{2m}\nabla^2\psi(\bold{r},t)+ m \Phi\psi(\bold{r},t)
\eeq
with a self-gravitation potential $\Phi$.
This equation could be obtained from the mean field approximation of a BEC Hamiltonian
 or the non-relativistic approximation
of SFDM action ~\cite{myhalo}.
For simplicity,  we consider the spherical symmetric case  with
\beq
   \Phi(r)=\int^{r}_0 dr'\frac{1}{r'^2}\int^{r'}_0 dr'' 4\pi r''^2
 (GmM |\psi(r)|^2+\rho_v),
\eeq
where $M$ is the mass of the halo,
and $\rho_v$ is the mass density of visible matter.
We do not consider a particle self-interaction term in this paper.

The Madelung representation ~\cite{2011PhRvD..84d3531C,2014ASSP...38..107S}
\beq
\psi(r,t)=\sqrt{A(r,t)}e^{iS(r,t)}
\label{madelung}
\eeq
is useful for studying
the cosmological structure formation in the fluid approach.
Here the amplitude $A$ and DM density have a relation $\rho=mA$.
Substituting Eq. (\ref{madelung}) in to GPE, one can obtain
a continuity equation
\beq
\frac{\partial \rho}{\partial t} + \nabla \cdot (\rho \textbf{v})=0
\label{continuity}
\eeq
and a modified Euler equation
\beq
\frac{\partial \textbf{v}}{\partial t} + (\textbf{v}\cdot \nabla)\textbf{v} +\nabla \Phi
+\frac{\nabla p}{\rho} -\frac{\nabla Q}{m} =0
\label{euler}
\eeq
with  a quantum potential $Q\equiv\frac{\hbar^2}{2m}\frac{\Delta \sqrt{\rho}}{\sqrt{\rho}}$,
a fluid velocity $\textbf{v}\equiv \nabla S/2m$, and  the pressure from a self-interaction pressure $p$ (if there is).
Here, $\Delta$ is the Laplacian.
The quantum pressure term
${\nabla Q}/{m}$ is the key difference between the CDM and the BEC DM.
Perturbing the equations (\ref{continuity}) and (\ref{euler})
 around $\rho=\bar\rho$, $\textbf{v}=0$, and $\Phi=0$ and then combining the two perturbed equations
 gives a differential equation for density
 perturbation $\delta\rho\equiv \rho-\bar{\rho}$,
 \beq
  \frac{\partial^2 \delta\rho}{\partial t^2}+\frac{\hbar^2}{4m^2}\nabla^2 (\nabla^2 \delta \rho)
  -c^2_s \nabla^2 \delta\rho - 4\pi G \bar{\rho}\delta\rho=0,
 \eeq
 where $c_s$ is the sound velocity from $p$, and $\bar{\rho}$ is the average background matter density (See, for example, Ref. \citealp{Suarez:2011yf}
 for details.).
 We have ignored the effect of the cosmic expansion in this equation for simplicity.
 We can rewrite this equation  into the Fourier transformed equation of the
  density contrast $\delta\equiv\delta \rho/\bar{\rho}=\delta_k e^{ik\cdot r}$ with a wave vector $k$,
  \beq
  \frac{d^2 \delta_k}{d t^2} +  \left[(c^2_q+c^2_s)k^2-4\pi G \bar{\rho} \right]\delta_k=0,
 \eeq
 where $c_q=\hbar k/2m$ is a quantum velocity.
 Note that the $k^4$ dependent term (the $c_q$ dependent term) came from the perturbation of the quantum pressure term.
 From this equation we can see that the BEC DM behaves like the CDM  for a small $k$ (for a large scale)
 while for a large $k$ (at a small scale) the quantum pressure disturbs the structure formation.
 If the self-interaction is negligible we can ignore the $c_s$ term.
Equating $c^2_q k^2$ with $4\pi G \bar{\rho}$ defines the time dependent quantum Jeans length scale ~\cite{fuzzy},
\beq
\label{lambdaQ}
\lambda_Q(z)= \frac{2\pi}{k}=\left(\frac{\pi^3 \hbar^2 }{m^2G\bar\rho(z)}\right)^{1/4}
\simeq 55593\left(\frac{\rho_b }{m_{22}^2\Omega_m h^2\bar\rho(z)}\right)^{1/4} pc,
\eeq
where the current matter density $\rho_b=2.775\times 10^{11} \Omega_m h^2~ M_\odot/ Mpc^{3}$, the  (dark + visible)  matter density parameter
$\Omega_m=0.315$ ~\cite{PDG-2014}, $h=0.673$
and $m_{22}=m/10^{-22}eV$.
The quantum Jeans mass can be defined as
\beq
\label{MJ}
M_J(z)=\frac{4\pi}{3} \bar{\rho}(z) \lambda_Q^3
=\frac{4}{3}
 \pi^{\frac{13}{4}}\left(\frac{\hbar}{G^{\frac{1}{2}}   m}\right)^{\frac{3}{2}} \bar{\rho}(z)^\frac{1}{4},
\eeq
which is the minimum mass of the DM structures at z.
Note that the only time dependent term in the righthand side is the average density.

Though $\lambda_Q$ is related to the minimum length scale of  DM dominated objects
~\cite{1985MNRAS.215..575K,Grasso:1990zg},
 $\lambda_Q$ alone does not determine the actual size of galaxies. Usually, $\lambda_Q > \xi > \lambda_c$.
  We need a governing equation for  stable configurations of the DM dominated
 objects.
To find the characteristic length $\xi$ we  study the ground state of the GPE.
In the BEC/SFDM model,
  $\xi\sim  \hbar /m \Delta v$ due to the uncertainty principle,
   where $\Delta v$ is the
 velocity dispersion of DM in a halo.  However, we have not been able to  derive   $\Delta v$ from any theory so far.
From Eq. (\ref{GPE}) the energy $E$ of the halo
 can be approximated as
\beq
E(\xi) \simeq\frac{\hbar^2}{2m \xi^2}+\int^{\xi}_0 dr'\frac{Gm}{r'^2}\int^{r'}_0 dr'' 4\pi r''^2
(\rho(r'')+\rho_{v}(r'') ),
\eeq
as a function of the halo length scale  $\xi$.
The  ground state   can be found by
extremizing it by $\xi$~\cite{Silverman:2002qx};
\beq
\label{dE}
\frac{dE(\xi)}{d\xi}\simeq -\frac{\hbar^2}{m \xi^{3}}+\frac{GM m}{\xi^2}=0,
\eeq
where,
$M\equiv \int^{\xi}_0 dr'' 4\pi r''^2
(\rho(r'')+\rho_{v}(r'') )$
 is the total mass within $\xi$.
Solving Eq. (\ref{dE}) gives  ~\cite{sin1,Silverman:2002qx}
\beq
\label{xi}
\xi=\frac{\hbar^2}{GMm^2}=\frac{c^2\lambda_c^2}{4\pi^2 GM}.
\eeq
 The quantum Jeans mass represents the smallest amount of the DM having enough self-gravity to overcome
the quantum velocity, so $M_J$ in Eq. (\ref{MJ}) can be identified to be $M$ of the smallest galaxies.
 Therefore, from Eq. (\ref{MJ}) the smallest galaxy formed at $z$ has a  size (the gravitational Bohr radius)
\beq
\label{xiz}
\xi(z)=\frac{\hbar^2}{G M_J(z)m^2}=\frac{3\hbar^{1/2}}{4\pi^{13/4} (G m^2 \bar{\rho}(z))^{1/4}}\propto \bar{\rho}(z)^{-1/4} ,
\eeq
which is a quantum mechanical relation absent in the CDM models. Therefore,
$M_J(z)$ and $\xi(z)$ represent
the time dependent mass and size of the smallest galaxies at the redshift $z$.
Recall that $M$ (and $M_J$) is the  total mass including DM and visible matter, which explains
the universal minimum mass of dwarf galaxies independent of visible matter fraction ~\cite{Lee:2008ux}.

Once we fix one of $\xi$ and $M_J$, the other  is fixed automatically.
In the previous works it was uncertain which one comes first, and
 several length scales including a thermal de Broglie wavelength or self-interaction scale
were considered for $\xi$. In Ref. \cite{Lee:2008ux} the thermal  de Broglie wavelength was proposed as $\xi$.
Now, considering the evolution of the density perturbation leading to  Eq. (\ref{MJ})
it is  reasonable to think that $\bar{\rho}(z)$ determines $M_J(z)$ first, and  $M_J(z)$ determines $\xi(z)$ in turn.
Therefore, the most natural length scale for galaxies is $\xi(z)$  given by Eq. (\ref{xiz}), if there is no self-interaction.

This simple argument leads to many interesting predictions.
Most of all, the smallest galaxies have the mass and the size determined by the epoch when the galaxies were formed.
Since $\bar{\rho}\propto a(z)^{-3}$, from Eq. (\ref{MJ}) we see
\beqa
\label{MJ2}
M_J(z)&=&M_J(0) a^{-3/4}(z)=M_J(0)(1+z)^{3/4}, \no
\xi(z)&=&\xi(0) a^{3/4}(z)=\xi(0)(1+z)^{-3/4},
\eeqa
which means that the minimum mass of galaxies decreases and
the size of the halos increases as the time flows. (See Fig. 1.)
\begin{figure}[htbp]
\includegraphics[width=0.6\textwidth]{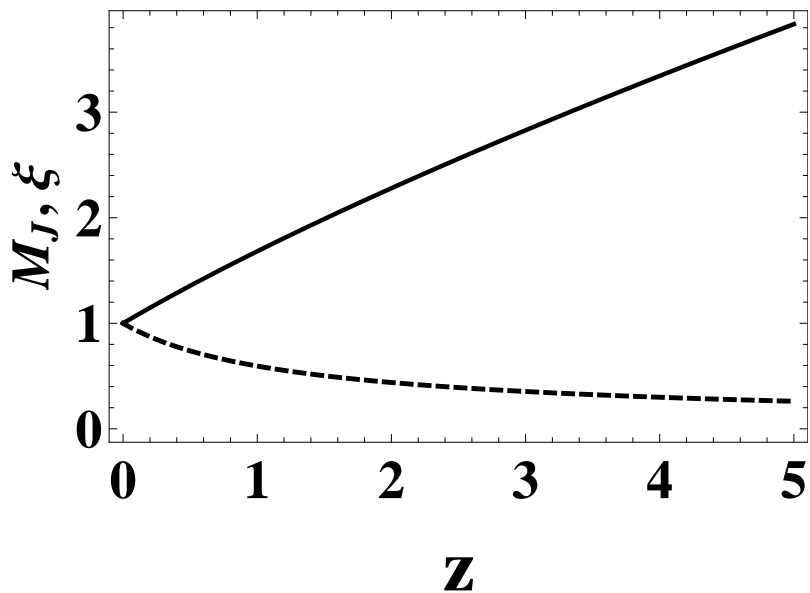}
\caption{ (Color online) The quantum Jeans mass $M_J(z)/M_J(0)$ (solid line) and the DM halo length scale $\xi(z)/\xi(0)$ (dashed line)
as functions of the redshift $z$.
For $m = 5\times 10^{-22}eV$ $M_J(0)=1.1\times 10^7 M_\odot$
and $\xi(0)=311.5 pc$.}
\end{figure}
Note that this does not mean the mass of typical galaxies decreases. Obviously,  galaxies can be  heavier
during  hierarchial merging processes.
 For the mass evolution above, we are considering only the smallest
DM dominated galaxies formed at  a given epoch.

The particle mass $m=O(10^{-22})eV$ is
required to solve the cusp problem
and the missing satellite problem ~\cite{fuzzy,Alcubierre:2001ea}.
For  $m=5\times  10^{-22}eV$, Eq. (\ref{MJ}) gives $M_J(0)=1.1\times 10^7 M_\odot$
and $\xi(0)=311.5 pc$. Interestingly, these values are  similar to
the minimum mass and the size of dSph galaxies nearby   obtained from astronomical data, respectively.
This is an interesting coincidence.

 The average mass density of the dwarf galaxies evolve as
\beq
\label{gz}
\rho_g(z)\sim\frac{M_J}{\xi(z)^3}\propto a^{-3}(z)=(1+z)^{3}.
\eeq
 Thus, another prediction of our model is that early dwarf galaxies are more compact than  present ones.
For example,  $M_J(z)=4.2\times 10^7 M_\odot$ and $\xi(z)=81.2 ~pc$ at $z=5$.
The high resolution numerical study with the  BEC/SFDM  \cite{Schive:2014hza,2014NatPh..10..496S} found
 that early galaxies  were compact, which also supports our model.
If these early compact dwarf galaxies are found in the sky, it could be another evidence for the BEC/SFDM.
 Interestingly, we already have similar galaxies.
The ultra-compact dwarf galaxies (UCD) are very compact
galaxies with high stellar populations. They are generally very old ($> 8 Gyr$), small ($< 100 pc$)
and they have mass $M\simeq 2-9\times 10^7 M_\odot$, which are similar to the predicted parameters of the early dwarf galaxies in our model.
Therefore, we conjecture that the UCD are remnants of these old dwarfs which have not experienced major mergers
since their formation, and have  kept an initial DM distribution.

\begin{figure}[htbp]
\includegraphics[width=0.6\textwidth]{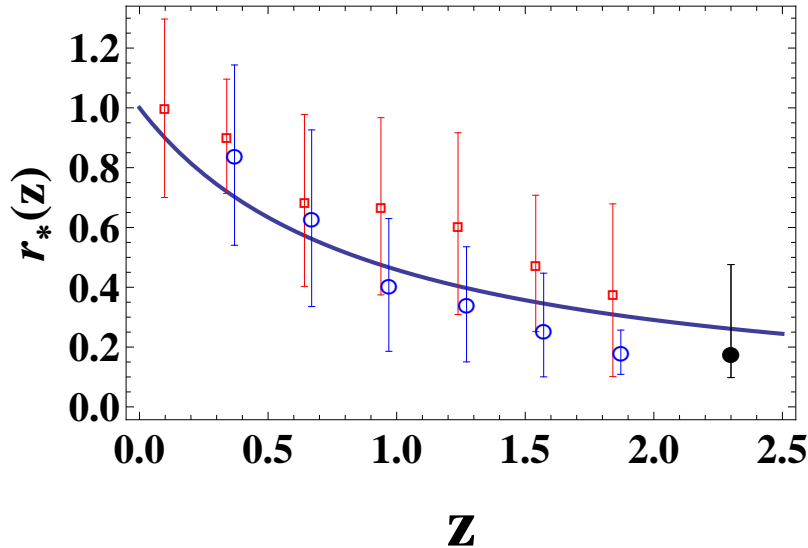}
\caption{ (Color online)
The observed size evolution $r(z)$ of the massive galaxies versus the redshift $z$
for spheroid-like galaxies (the red squares), disc-like galaxies (the blue circles) (Data from \cite{Trujillo21112007}) , and
a typical compact galaxy (the black dot) (Data from \cite{2009Natur.460..717V}).
The line represents the predicted
size evolution of the visible parts of the galaxies  $r_{*}(z)=1/(1+z)^{9/8}$ in our model,
which agrees with the observational data. We have set $r_*(0)=1$.}
 \label{r}
\end{figure}

Owing to the scaling properties of the BEC DM halos, we can assume that this size evolution happens also to  the most massive galaxies.
The visible part of a galaxy seems to scale as $r_*(z)\propto \xi(z)^{3/2}$ \cite{Lee:2008ux},
 so we expect the size of the very massive and compact galaxies evolves as
\beq
\label{rstar}
\frac{r_*(z)}{r_*(0)}= a^{9/8}(z)=(1+z)^{-1.125},
\eeq
which is shown in Fig. 2.
The predicted
size of the visible parts of the galaxies
turns out to be similar to the average value of disk-like galaxies and spheroid-like galaxies, and to be
  in a good agreement  with the observational data.
Furthermore,
HST /WFC3 IR images  were analyzed to study evolution of quiescent galaxies \cite{Morishita:2014xua}, and it was found that
the size evolution of the massive galaxies follows $r_*\propto  (1 + z)^{-\alpha}$ with $\alpha = 1.06 \pm 0.19$, which
is also consistent with our prediction.
(Our prediction for $\alpha$ value is different from that of the previous work~\cite{Lee:2008ux} because of the choice of $\xi$.)

One can see the ratio $\lambda_Q(z)/\xi(z)=4\pi^4/3\simeq 129.8$
is a constant independent of the time
from Eq. (\ref{lambdaQ}) and Eq. (\ref{xiz}). This number is
the contraction factor during the DM collapse to form a halo. The quantum Jeans length indeed decides the length scale
but the actual galaxy scale is much smaller by this factor.

On the other hand, the  ratio ${\lambda_c}/{\xi}$ represents how relativistic a given halo is. Using Eq. (\ref{xiz}) we find
\beq
\frac{\lambda_c}{\xi}=\frac{4\pi^2 GM_J}{c^2\lambda_c}=
\frac{2^{9/4} \pi^4 \hbar^{1/2}\Omega_m^{1/4}}{3^{3/4} c}\sqrt{\frac{H}{m}}>\frac{4\pi^2 GM_J}{c^2\xi},
\eeq
which is proportional to the ratio of gravitational potential energy  and the self energy of the halo DM particles.
Here, $H\simeq 10^{-33}eV$ is the Hubble parameter and we have used the Friedmann equation $H^2=8\pi G \rho_c/3$,
where $\bar{\rho}=\Omega_m \rho_c$.
From this we can understand why galaxies are non-relativistic ($\lambda_c/\xi\ll 1$).
It is due to the  small $M_J$,  or
the small matter density $\bar{\rho}$ at the epoch of  galaxy formations (See Eq. (\ref{MJ})).
More fundamentally, this is caused by the small ratio of the Hubble parameter to the mass $m$, i.e.,
$\sqrt{H/m}\sim 10^{-6}$.
From Eq. (\ref{xiz}) we  obtain
\beq
{\xi}=\frac{3^{3/4} \hbar^{1/2}}{2^{5/4} \pi^3 \Omega_m^{1/4} \sqrt{H m}}\simeq  0.00656 \sqrt{\frac{m}{H}} \lambda_c.
\eeq
It is now clear that the rotation velocity $V_{rot}\simeq \Delta v$ of typical dwarf galaxies is
\beq
V_{rot}=\sqrt{\frac{GM_J}{\xi}}=
\frac{2^{5/4} \pi^3 \hbar^{1/2}\Omega_m^{1/4}}{3^{3/4} }\sqrt{\frac{H}{m}}  \simeq 5\times 10^{-5} c,
\eeq
which is similar to the observed value $V_{rot}=O(10) km/s$.
Thus, the characteristic  rotation velocity of galaxies has a quantum origin
and is  $O(\sqrt{H/m})c$, which is somewhat surprising.

In summary,  the  BEC/SFDM can explain not only the  problems of the CDM but also
 the  observed minimum size and the mass of  galaxies with a few parameters.
The background matter density (or equivalently the Hubble parameter) and $m$ decide the coherence length of the
 BEC halos, and this length in turn decides the rotation velocity. In this model the minimum size of galaxies increases, while
the minimum mass of the  galaxies decrease as the universe evolves.
This scenario seems to explain the observed size evolution of massive galaxies and gives a new hint to
the origin of the UCD.

\acknowledgments
This work is supported by the sabbatical year program of Jungwon university in 2015.

\vskip 5.4mm
%

\end{document}